\documentclass[11pt]{article}

\usepackage{hyperref}
\hypersetup{
    colorlinks=true,
    linkcolor=blue,
    filecolor=magenta,      
    urlcolor=blue,
}

\usepackage{color}
\usepackage{latexsym}
\usepackage{epsfig,graphics}
\usepackage{amsmath,amssymb,amsthm}
\usepackage{graphics,epsfig,calc}
\usepackage{bm}

\DeclareMathAlphabet\mathbfcal{OMS}{cmsy}{b}{n}

\usepackage{upref}
\usepackage{setspace}
 \newcommand{\beqn}{\begin{eqnarray}}
 \newcommand{\eeqn}{\end{eqnarray}}
 \newcommand{\be}{\begin{equation}}
 \newcommand{\ee}{\end{equation}}
 \newcommand{\ba}{\begin{array}}
 \newcommand{\ea}{\end{array}}

 \newcommand{\pa}{\partial}
 
  \newcommand{\ci}{\cite}
 \newcommand{\ds}{\displaystyle}
 \newcommand{\la}{\label}

  \newcommand{\eff}{{\rm eff}}

 \newcommand{\fr}{\frac}
 
 \newcommand{\we}{{\rm \wedge}}

\newcommand{\toLd}{\stackrel{L^2}\longrightarrow} 
\newcommand{\toH}{\stackrel{H^1}\longrightarrow}

\newcommand{\ti}{\tilde}

\newcommand{\cF}{{\cal F}}

\newcommand{\bA}{{\bf A}}

\newcommand{\cD}{{\mathbb D}}

\newcommand{\cB}{{\cal B}}
\newcommand{\cE}{{\cal E}}

\newcommand{\cS}{{\cal S}}

\newcommand{\bH}{{\bf H}}

\newcommand{\bj}{{\bf j}}
\newcommand{\bJ}{{\bf J}}
\newcommand{\cJ}{{\cal J}}
\newcommand{\cbJ}{{\mathbb J}}

\newcommand{\bbl}{{\bf l}}
\newcommand{\cM}{{\mathbb{M}}}

\newcommand{\cP}{{\cal P}}
\newcommand{\cQ}{{\cal Q}}

\newcommand{\bS}{{\bf S}}

\newcommand{\aT}{{\mathbb{T}}}

\newcommand{\cV}{{\mathbb{V}}}

\newcommand{\bY}{{\bf Y}}

\newcommand{\cY}{{\mathbb{Y}}}
\newcommand{\cZ}{{\mathbb{Z}}}

\newcommand{\bZ}{{\bf Z}}

\newcommand{\ve}{\varepsilon}

\newcommand{\De}{\Delta}
\newcommand{\de}{\delta}
\newcommand{\dv}{{\rm div\5}}
\newcommand{\curl}{{\rm curl\5}}

\newcommand{\al}{\alpha}
\newcommand{\ga}{\gamma}

\newcommand{\vka}{{\varkappa}}

\newcommand{\om}{\omega}

\newcommand{\na}{\nabla}

\newcommand{\bal}{{\bm\al}}
\newcommand{\bbeta}{{\bm\beta}}
\newcommand{\bga}{{\bm\ga}}
\newcommand{\bxi}{{\bm\xi}}

\newcommand{\bPi}{{\bm\Pi}}

\newcommand{\Si}{\Sigma}
\newcommand{\lam}{\lambda}
\newcommand{\Lam}{\Lambda}

\newcommand{\5}{{\hspace{0.5mm}}}

\newcommand{\N}{\mathbb{N}}
\newcommand{\R}{\mathbb{R}}

\newcommand{\C}{\mathbb{C}}

\newtheorem{theorem}{Theorem}[section]

\renewcommand{\thetheorem}{\arabic{section}.\arabic{theorem}}
\newtheorem{defin}[theorem]{Definition}

\newtheorem{lemma}[theorem]{Lemma}
\newtheorem{remark}[theorem]{Remark}

\newtheorem{cor}[theorem]{Corollary}
\newtheorem{pro}[theorem]{Proposition}
\newcommand{\bp}{\begin{pro}}
\newcommand{\ep}{\end{pro}}

\newcommand{\bt}{\begin{theorem}}
\newcommand{\et}{\end{theorem}}

\newcommand{\bl}{\begin{lemma}}
\newcommand{\el}{\end{lemma}}

\newcommand{\const}{\mathop{\rm const}\nolimits}

\newcommand{\bce}{\begin{center}}
\newcommand{\ece}{\end{center}}

\newcommand{\bpr}{\begin{proof}}
\newcommand{\epr}{\end{proof}}

\newcommand{\br}{\begin{remark}}
\newcommand{\er}{\end{remark}}

\newcommand{\bd}{\begin{defin}}
\newcommand{\ed}{\end{defin}}

\newcommand{\bc}{\begin{cor}}
\newcommand{\ec}{\end{cor}}

\begin{document}

\bce
{\Large\bf  On stability  of solitons for  3D
Maxwell--Lorentz 
\medskip

system  with spinning particle}
 \bigskip \medskip

{\Large A. I. Komech
 \footnote{ 
 Supported partly by Austrian Science Fund (FWF) PAT 3476224}
 }
 \\ 
{\it
\centerline{Institute of Mathematics, BOKU University, Vienna
}

\centerline{alexander.komech@gmail.com}

\medskip

 {\large E.A. Kopylova}
 \footnote{ 
 Supported partly by Austrian Science Fund (FWF) P34177
 }
 \\
{\it Faculty of Mathematics,   Vienna University
  }
 \centerline{elena.kopylova@univie.ac.at}

 }
 
 \ece

\begin{abstract}
We consider stability of solitons of 3D Maxwell--Lorentz system
with extended charged spinning particle. 
The solitons are  solutions which correspond to a particle
 moving with a constant velocity $v\in\R^3$ with $|v|<1$
and rotating with a constant angular velocity $\om\in R^3$. 

 Our main results are 
the   orbital stability of moving solitons  with $\om=0$
 and  a {\it linear} orbital stability of 
  rotating solitons with $v=0$.

   The  Hamilton--Poisson structure of the Maxwell--Lorentz system
    is degenerate and 
   admits the Casimir invariants.
   We construct the  Lyapunov function as  a linear combination of the 
    Hamiltonian with a suitable Casimir invariant.
    The key point is
      a lower bound for this function.
    The proof of the bound in the case $\om\ne 0$
    relies on angular momentum conservation
    and
  suitable spectral arguments including the Heinz--L\"owner--Kato inequality and closed graph theorem.

\end{abstract}

\newpage
\tableofcontents

\section{Introduction}

The paper addresses
the stability of solitons of 3D Maxwell--Lorentz system
with an extended charged spinning particle. 
We choose the units such that the speed
of light is $c=1$.
The  equations  read as follows
(see \ci{N1964,S2004}):
\begin{equation} \la{mls3}
\left\{\begin{array}{rcl}
\dot E(x,t)&=&\curl B(x,t)-w(x,t)\rho(x-q(t)),\qquad 
\dv E(x,t)=\rho (x-q(t))
\\
\dot B(x,t)&=& -\, \curl E(x,t),\qquad\qquad\qquad\quad \quad \quad\quad
\dv B(x,t)=0
\\
m\dot q(t)&=&p(t),\quad
\dot p(t)=\langle E(x,t)+w(x,t)\we B(x,t),\rho(x-q(t)) \rangle
\\
 I\dot \om(t)&=&\langle (x-q(t))\we
\big[E(x,t)+w(x,t)\we B(x,t)\big],
\rho(x-q(t))\rangle
\end{array}\right|, 
\end{equation}
where
$\rho(x-q(t))$ is the charge distribution of the extended 
particle centered at the point $q(t)\in\R^3$, 
$m>0$  is the mass of the particle and $I>0$ is its moment of inertia,
$\om(t)$ is the angular velocity 
of the particle rotation and $w(x,t):=\dot q(t)+\om(t)\we(x\!-\!q(t))$ is the velocity field. 
{ The brackets $\langle\cdot,\cdot\rangle$ here and below 
denote the 
integral over $x\in\R^3$ or  inner product in the Hilbert spaces
$L^2:=L^2(\R^3)\otimes\C^n$ with suitable $n\ge 1$.
}
This model
was introduced by  Abraham in 1903--1905 (see \ci{A1903, A1905})
for the description of the classical extended electron coupled to its own Maxwell field.  
 The detailed account on the genesis and  early investigations 
 of the system
 by Dirac,  Poincar\'e, Sommerfeld, and others
 can be found in \ci[Chapter 3]{S2004}.
\smallskip
 
 The first results on the corresponding long-time behavior 
 for the system (\ref{mls3})
 were obtained in the 1990s.
This system
plays a crucial role in a rigorous analysis of radiation by moving particles; see
  \ci{KK2022,S2004}.
The  mathematical analysis of  this system  
is useful in connection to the related problems  of 
nonrelativistic QED. 
In particular, the similarity in the renormalization of mass
was pointed out
by Hiroshima and Spohn \ci{HS2005}.

\smallskip

We assume that the charge density
$\rho(x)$ is smooth, spherically-symmetric, and not identically zero:
\be\la{rosym}
\rho\in C_0^\infty(\R^3),\qquad\rho(x)=\rho_1(|x|);\qquad \rho(x)=0\,\,\,\,{\rm for}\,\,|x|\ge R_\rho;\qquad \rho(x)\not\equiv 0.
\ee
One of the basic peculiarities of 
the system (\ref{mls3}) is its  invariance under
 the Euclidean group generated by 
 translations, rotations and reflections of the space $\R^3$.
 Moreover, the system
 admits solitons $\cS_{v,\om}$
 discovered by Spohn \ci{S2004}.
  The solitons are solutions
 moving with constant speed 
$v\in\R^3$ 
and rotating with constant angular velocity $\om\in\R^3$.
The solitons are trajectories of 
one-parametric subgroups of the 
Euclidean
symmetry group. 
As is pointed out in \ci{S2004},
for  $\rho(x)\not\equiv 0$,
the solitons with finite energy exist only 
for $(v,\om)\in\Si$, where
\be\la{omv}
\Si=\{(v,\om)\in\R^3\times\R^3:|v|<1\,\,{\rm and}\,\, {\rm either}\,\, \om\Vert v\,\, {\rm or}\,\,\,\om\bot v\}.
\ee 
In Appendix B,
we justify
the restrictions (\ref{omv}) on the parameters of solitons 
and calculated  effective moment of inertia of solitons (\ref{Ieff}).

\br\la{r0}
\rm

In the case $\rho(x)\equiv 0$ we have $\ddot q(t)=\dot\om(t)=0$ and
the Maxwell equations do not depend on the motion of the particle, so the energy of the Maxwell field is conserved, and so are the particle momentum
and angular momentum.
The  solitons read as $(E(x,t),B(x,t), q(t),p(t),\om(t))=(0,0,vt+q_0, mv,\om)$ with arbitrary $q_0, v,\om\in\R^3$,  and  they are obviously orbitally stable
 by the conservation laws; see Definition \ref{dsta}.

\er

\noindent
Our main results are i)
the  orbital stability of solitons $\cS_{v,0}$ with $|v|<1$
 and ii) the {\it linear} orbital stability
 of solitons $\cS_{0,\om}$ with  $\om\in\R^3$ 
 in the class of solutions with zero total momentum.
In particular, the total momentum is zero for solutions
with ``symmetric"  initial data:
\be\la{inid}
q(0)= p(0)=0,\qquad E(-x,0)\equiv-E(x,0),\qquad B(-x,0)\equiv B(x,0).
\ee 
Our approach essentially relies on the Hamilton--Poisson structure of the 
system (\ref{mls3}) which was constructed in \ci{KK2023_LH}.
The structure is degenerate, so the system admits the Casimir invariants,
and the methods of \ci{GSS} are not applicable to the system.

    The proof of the orbital stability of solitons $\cS_{v,0}$ relies on the reduction  to the comoving frame and lower bound
for the Hamiltonian as the Lyapunov function.
The Hamilton functional of the reduced system depends on the parameter $P$ which is the conserved  total momentum. Thus, we restrict the Hamiltonian to the states with the fixed value of $P$. 

To study the stability of solitons $\cS_{0,\om}$, we employ conservation of 
 angular momentum $J$. The angular momentum is a functional on the phase space
 of the reduced system only for the total momentum $P=0$. So, we 
consider the stability in the class of solutions with $P=0$.
In particular, 
 all solutions with initial states (\ref{inid}) have zero total momentum.
 
The {\it linear orbital stability} of the
solitons refers to 
the stability of the 
{\it linearized reduced dynamics}
in the tangent space to the manifold of states with constant angular momentum.
The manifold is not well-defined, but the tangent space is;
see Section \ref{som}.
So, the nonlinear stability problem on the manifold is not correct,
as opposed to
the linearized problem on the tangent space.

  To prove the linear orbital stability,  
  we 
    construct the Lyapunov function $L$ 
    for the linearized equations
    as the sum of the 
    quadratic part of the  
     Hamiltonian with the Casimir functional. 
     Such a strategy is an
     analog of the energy--Casimir method
     \ci{HMRW1985,M2002}; see Remark \ref{rec}.
     We prove the conservation of $L$ in the  linearized dynamics
     and
     establish a suitable lower bound for  $L$.
      A lower bound provides a priori estimate 
 and existence of global solutions, and at the same time
      the stability of the  linearized dynamics.
      The bound is  proved 
by suitable spectral arguments including the Kato--L\"owner--Heinz inequality from the theory
of interpolation \ci{K1952} and the closed graph theorem.   
     Further comments on our methods and open questions can be found in
     Appendix
     \ref{sLJ}.
     
  \smallskip

Let us comment on related work.
{ The mathematical theory of Maxwell--Lorentz system (\ref{mls3}) started  after} 
 Nodvick's paper \ci{N1964}, where the Hamilton least action principle 
 and conservation laws
have been established
 in the Euler angles representation.
 The coordinate-free proof of the conservation laws 
was given by Kiessling \ci{K1999}. 
   The coordinate-free proof of the Hamilton principle
was given in
\ci{IKS2015} using the technique of \ci{AKN1997,P1901} which relies on the 
Lie algebras and the Poincar\'e equation.
 This technique was developed 
in \ci{IKS2017}, where the general theory of invariants was constructed
for the Poincar\'e equations on manifolds and applied to the system (\ref{mls3}). 
    
      In \ci{BG1993,IKM2004,IKS2011,KK2023jmp,KS1998,KS2000,KSK1997}, the stability of solitons was studied
     for the  Maxwell--Lorentz system  (\ref{mls3}) 
     without spinning 
(the first three lines with $w(t)\equiv \dot q(t)$).
   The results \ci{IKM2004,KS2000}
 provide the first rigorous proof of the {\it radiation damping}
 in classical electrodynamics.
 For the survey, see \ci{S2004} and \ci{KK2020,KK2022}.
 The adiabatic effective dynamics of solitons for the 3D system was proved 
  in  \ci{KS2000ad}.

{ For the Maxwell--Lorentz system with spinning particle
 at rest
 ($q(t)\equiv 0$),
the global convergence to rotating solitons 
 was  established in \ci{IKS2004} and  \ci{K2010}.
 In \ci{IKS2004},
 the convergence is proved
 in the case of sufficiently small charge density $\rho$.}
 In \ci{K2010}, Kunze extended the result to 
  the same system without the smallness condition.
 This condition  is replaced by the assumption that
  the mass of the particle does not belong to a discrete set  of resonances.
 
  The  {  stability of solitons} for the 3D Maxwell--Lorentz system (\ref{mls3})   with moving and spinning particle was not considered previously.

\setcounter{equation}{0}
\section{The Maxwell potentials and the well-posedness}
As is well-known \ci{Jackson}, the second line of the system (\ref{mls3}) implies
the following expressions of the Maxwell fields in terms of the potentials
$A(x,t)=(A_1(x,t),A_2(x,t),A_3(x,t))$ and $\Phi(x,t)$:
in the Coulomb gauge,
\be\la{AA}
B(x,t)=\curl A(x,t),\qquad E(x,t)=-\dot A(x,t)-\na \Phi(x,t),\qquad \dv A(x,t)=0.
\ee
Then the  first line  of the system (\ref{mls3}) becomes
\be\la{APD}
-\ddot A(x,t)-\na\dot\Phi(x,t)=-\De A(x,t)-
w(x,t) \rho(x-q(t)),\qquad \De\Phi(x,t)=\rho(x-q(t)).
\ee
Hence, $\Phi(x,t)$ is the Coulomb potential $\Phi(x,t)=-\ds\fr1{4\pi}\int\fr{\rho(y-q(t))}{|x-y|}dy$.
The gradient $\na\dot\Phi(x,t)$
in the first equation in (\ref{APD})
 can be eliminated
by the application of the    
orthogonal projection $\cP$ in $L^2$ onto the 
divergence-free vector fields. 
Substituting the expressions (\ref{AA}) into
the system  (\ref{mls3}) 
and using the notation $j(x,t):=
\cP([\dot q(t)+\om(t)\we (x-q(t))] \rho(x-q(t)))$ for the current density,
we obtain the system
\be\la{2ml2}
\left\{\ba{rcl}
\ddot A(x,t)\!\!&\!\!=\!\!&\!\!\De A(x,t)+ j(x,t)
\\
m\ddot q(t)\!\!&\!\!=\!\!&\!\!\langle-\dot A(x,t)\rho(x\!-\!q(t))+
j(x,t)\we\curl A(x,t)  \rangle
\\
 I\dot \om(t)\!\!&\!\!=\!\!&\!\!\langle (x\!-\!q(t))\we 
 [-\dot A(x,t)\rho(x\!-\!q(t))+
j(x,t)\we\curl A(x,t) ] \rangle
 %
 %
\ea\right|.
\ee
 In the last two equations, the terms with $\Phi(x,t)$ cancel since
$$
 \langle\na\Phi(x,t),\rho(x\!-\!q)\rangle=
 \langle -ik\frac{\hat\rho(k)}{k^2},\hat\rho(k)\rangle=0,\quad 
 \langle  (x\!-\!q)\na\Phi(x,t),\rho(x\!-\!q)\rangle=-\langle \na \we k\frac{\hat\rho(k)}{k^2},\hat\rho(k)\rangle=0.
$$
Indeed,
the first equality holds
since  the function $\hat\rho(k)$ is even.
The second equality follows from the rotation invariance (\ref{rosym}).


Denote the  spaces of real functions 
$C^k=C^k(\R^3)\otimes\R^3$   with $k=0,1,\ldots$,
the Sobolev spaces
$H^s=H^s(\R^3)\otimes\R^3$ with $s\in\R$,
and  $\dot H^1=\dot H^1(\R^3)\otimes\R^3$
being the completion of $C_0^\infty:=C_0^\infty(\R^3)\otimes\R^3$ with the norm
$\Vert A\Vert_{\dot H^1}=\Vert\na A\Vert_{L^2}$. 

\bd Denote i)
$\cF^0\!:=\!\{A\!\in\! L^2\!: \dv A(x)\!\equiv\! 0\}$ the Hilbert space
endowed with the norm of $L^2$, and ii) $\dot\cF^1\!:=\!\{A\!\in\! \dot H^1\!: \dv A(x)\!\equiv \!0\}$ the Hilbert space which is the completion of $\{A\in C_0^\infty:\dv A(x)\equiv0\}$ with respect to the norm $\Vert \na A\Vert_{L^2}$.
\ed

\br\la{rde}
\rm
The space $C_0^\infty\cap\cF^0 $ is dense in $\cF^0$ (see Appendix \ref{aC}).
\er

Denote the Hilbert spaces
\be\la{cWH}
\cY:=\dot\cF^1\oplus\cF^0\oplus \R^3\oplus \R^3\oplus \R^3,\qquad
\cV=L^2\oplus H^{-1}\oplus\R^3\oplus \R^3\oplus \R^3.
\ee
For $n=1,2,3$
let us denote
\be\la{na*}
\langle \dot A(x), \na_* A(x)\rangle_n\!=\!\langle \dot A(x), \na_n A(x)\rangle,
\quad\langle ((x\!-\!q)\we \na)_*A(x),\dot A(x)\rangle_n\!=\!\langle ((x\!-\!q)\we \na)_nA(x),\dot A(x)\rangle.
\ee
The notations are suitable versions of 
the basic operation {\rm \ci[(10)]{A1966}} from the 
theory of coadjoint representation 
for the groups of translations and rotations, respectively.
Denote by $p$ and $\pi$ 
the momentum and angular momentum
of the particle in the Maxwell field,
\be\la{deo2}
p
=m\dot q+\langle A(q+y),\rho(y)\rangle,
\qquad
\pi
=I\om+\langle y\we A(q+y),\rho(y)\rangle.
\ee

\bp\la{pwpA}
{\rm i)} For any initial state $Y(0)=(A(x,0),\dot A(x,0),q(0),p(0),\om(0))\in\cY$, the system {\rm (\ref{2ml2})} admits a unique
solution 
\be\la{wpY}
Y(t)=(A(x,t),\dot A(x,t),q(t),p(t),\om(t))\in C(\R,\cY)\cap C^1(\R,\cV).
\ee
{\rm ii)} The map $W(t):Y(0)\mapsto Y(t)$
 is continuous in $\cY$ for every $t\in\R$, and
the energy and momentum are conserved{\rm :}
\beqn\la{ecoA}
\cE(t)&=&\fr12\int[\dot A^2(x,t)+|\curl A(x,t)|^2]dx+\fr 12m\dot q^2(t)
+\fr12I \om^2(t)=\const,\qquad t\in\R,
\\
P(t)&=&-\langle \dot A(x,t), \na_* A(x,t)\rangle+ p(t)+\langle\rho(x-q(t)), A(x,t)\rangle=\const,\qquad\,\,\, t\in\R.
\la{mcoA}
\eeqn

{\rm iii)} Let for  $|\al|\le 3$ and $|\beta|\le 2$,
\be\la{AP0}
A(x,0)\!\in \!C^3\!\cap \!\dot\cF^1,\quad \dot A(x,0)\!\in\! C^2\!\cap\!\cF^0;
\quad
|\pa_x^\al A(x,0)|+|\pa_x^\beta \dot A(x,0)|\le  C(1+|x|)^{-3},\quad x\in\R^3.
\ee
Then $\pa_x^\al \pa_t^l  A(x,t)\in C^{2-(|\al|+l)}(\R^3\times [0,T])\!\otimes\!\R^3$  for any $T>0$, $l=0,1,2$, and 
$|\al|+l\le 2$, and 
\be\la{AP1}
\left\{\ba{rcll}
|\pa_t^l A(x,t)|&\le& C(1+|x|)^{-2}, &l\le2
\\\\
|\pa_x^\al\pa_t^l  A(x,t)|&\le& C(1+|x|)^{-3},&\al\ne 0,\,\,|\al|+l\le 2
\ea\right|,\qquad x\in\R^3,\,\,\, |t|<T.
\ee
For such solutions,
\be\la{amcoA}
J(t)=\langle A(x,t)\we \dot A(x,t)\rangle-
\langle ((x-q(t))\we \na)_*A(x,t),\dot A(x,t)\rangle+q(t)\we P+\pi(t)=\const,\,\,t\in\R.
\ee
\ep

The items i) and ii) were proved in \ci{KK2023_LH}, and 
 (\ref{amcoA}) is proved in 
 \ci{K1999} by a partial integration.
 The bounds (\ref{AP0}), (\ref{AP1}) justify the calculations.
Let us comment on the proof of (\ref{AP1}).
In the first equation in (\ref{2ml2}), the current
reads
$$
j(x,t):=j_0(x,t)+\om(t)\we (x-q(t))\rho(x-q(t)), \qquad j_0(x, t)=\cP([\dot q(t)\rho(\cdot-q(t))].
$$
In the Fourier transform, $\hat j_0(k,t)=e^{iq(t)}
[\dot q(t)-\fr{k (\dot q(t)\cdot k)}{k^2}]\hat\rho(k) $. 
Hence, $|\pa_k^\beta [k^\al\hat j_0(k,t)]|\sim |k|^{|\al|-|\beta|}$ as $|k|\to 0$. Therefore,
$\pa_k^\beta [k^\al\hat j_0(k,t)]\in L^1(\R^3)$ for $|\beta|\le 2+|\al|$, so
$$
|\pa_x^{\al} j_0(x, t)|\le C_\al(1+|x|)^{-2-|\al|},\quad x\in\R^3,\qquad\forall \al.
$$
Now the decay (\ref{AP1}) follows from the Kirchhoff formula for solutions
to the wave equation.

\setcounter{equation}{0}
\section{The Hamilton--Poisson structure}
In this section  we recall
the Hamilton--Poisson structure   of the system (\ref{2ml2})
which has been
constructed in  \ci{KK2023_LH}. The construction
relies on 
the Hamilton least action principle  for the  system (\ref{2ml2}) 
established
in \ci{IKS2015,N1964}.
Using this principle and the Lie--Poincar\'e technique \ci{A1966,H2009},
we have shown in
  \ci{KK2023_LH}  that  
  the system (\ref{2ml2})
 admits the following representation:
\be\la{2ml2rd}
\dot A=\Pi,\,\,\dot \Pi=\De A+j;\,\,\,
m\dot q=p-\langle A(q+y),\rho(y)\rangle, \,\, \dot p=\langle [ \dot q+\om\we y]\rho(y),\na_* A(q+y)\rangle;\,\,\,
 \dot \pi=\om\we\pi.
\ee
Let us rewrite
the energy (\ref{ecoA})
as the Hamiltonian functional:
 \beqn\la{Ham}
\!\!&\!\!\!\!&\!\!H(A,\Pi,q,p,\pi)=
\fr12\langle\Pi,\Pi\rangle+
\fr12\langle\curl A,\curl A\rangle+ \fr12m\dot q^2+\fr12I\om^2
\nonumber\\
\!\!&\!\!=\!\!&\!\!
\fr12\!\int[|\Pi(x)|^2\!+\!|\curl A(x)|^2]dx\!+\!\fr1{2m} [p\!-\!\langle A(x),\rho(x\!-\!q)\rangle]^2
\!+\!\fr1{2I}[\pi\!-\!\langle (x\!-\!q)\we A(x),\rho(x\!-\!q)\rangle]^2.
\qquad
\eeqn
The Hamiltonian is well-defined and Fr\'echet-differentiable
on the Hilbert phase space  $\cY$ defined in (\ref{cWH}).
The system (\ref{2ml2rd})
admits the 
Hamilton--Poisson 
 representation
\be\la{Hf}
\dot Y=\cJ(Y) D H(Y),\qquad Y=(A,\Pi,q,p,\pi),\qquad
\cJ(Y)=\left(
\ba{ccccc}
0&1&0&0&0\\
-1&0&0&0&0\\
0&0&0&1&0\\
0&0&-1&0&0\\
0&0&0&0&-\pi\we
\ea
\right).
\ee
Note that the operator $\cJ(Y)$ is not invertible for all $ Y$. Accordingly,
the system admits the {\it Casimir invariants} 
\be\la{Cas}
C(Y)=\pi^2,\qquad Y=(A,\Pi,q,p,\pi).
\ee

\setcounter{equation}{0}
\section{The canonical transformation to the comoving frame}
 
Denote the fields in the comoving frame
 \be\la{fcom}
 \bA(y,t):=A(q(t)+y,t), \qquad \bPi(y,t):=\Pi(q(t)+y,t).
 \ee
 We  are going to change the variables in the system (\ref{Hf})
 via the map
 \be\la{T}
T: (A,\Pi,q,p,\pi)\mapsto (\bA,\bPi,q,P,\pi),\qquad 
P=p-\langle \bPi, \na_* \bA\rangle.
\ee
Here $P$ is the conserved total momentum (\ref{mcoA}).
Hence, we can define
\be\la{Ham2d}
\bH(\bA,\bPi,q,P,\pi):=H(A,\Pi,q,p,\pi),\qquad p=P+\langle \bPi, \na_* \bA\rangle.
\ee

 \bl
The system {\rm (\ref{Hf})} in the new variables is equivalent to
a similar system with the Hamiltonian $\bH$:
\be\la{Hf2}
\dot \bY(t)=\cJ(\bY(t)) D \bH(\bY(t)),\qquad \bY(t)=(\bA(t),\bPi(t),q(t),P(t),\pi(t)).
\ee 
 
 \el
\bpr First, the map (\ref{T})
is canonical, i.e., it leaves the canonical form unchanged: 
\be\la{Tca}
\langle\bPi,\dot \bA\rangle+P\cdot\dot q+\pi\cdot\om=\langle\Pi,\dot A\rangle+p\cdot\dot q+\pi\cdot\om.
 \ee
 Indeed,
 $\dot \bA(y,t)=\dot A(q(t)+y,t)+\dot q(t)\cdot\na A(q(t)+y,t)$ by (\ref{fcom}).
 Hence, the identity (\ref{Tca}) reduces to
 $
\langle\Pi(q+y),\dot q\cdot\na A(q+y)\rangle+
P\cdot\dot q=p\cdot\dot q,
 $
 which holds by (\ref{T}).
Second, 
the identities (\ref{Tca}) and (\ref{Ham2d}) imply that the Lagrangians 
corresponding to the Hamiltonians $\bH$ and $H$ are related via the map $T$:
$
\bbl(\bA,q,\dot\bA,\dot q,\om)\equiv l(A,q,\dot A,\dot q,\om).
$
Therefore, the Lagrangian actions also coincide:
$
\int_a^b\bbl(\bA,q,\dot\bA,\dot q,\om)dt=\int_a^b l(A,q,\dot A,\dot q,\om)dt.
$
Hence, the solutions of the systems (\ref{Hf}) and (\ref{Hf2})
 also are related by this transformation since they are critical trajectories of the action.
\epr

\br\rm
Similar canonical transformations have been applied in \ci{BG1993,IKM2004}
to the systems of type (\ref{mls3}) without spinning and in \ci{KS1998} to the system
of  the particle coupled to the scalar field.

\er

\setcounter{equation}{0}
\section{The reduced system}

The Hamiltonian (\ref{Ham2d}) does not depend
on the variable $q$. Hence, the equations for $q$ and $P$
in  the system (\ref{Hf2})
 reduce to
\be\la{doq}
\dot q=D_P\bH,\qquad \dot P=0,
\ee
 which correspond to the conservation
of the conjugate momenta $P$.
As a result, 
the system
 (\ref{Hf2}) reduces to the  family of the 
Hamiltonian systems with the parameter $P\in\R^3$: for 
\be\la{Hf3}
\dot \bZ(t)=\cbJ(\bZ(t)) D \bH_P(\bZ(t)),\qquad \bZ(t)=(\bA(t),\bPi(t),\pi(t)),\qquad
\cbJ (\bZ)=\left(
\ba{ccccc}
0&1&0\\
-1&0&0\\
0&0&-\pi\we
\ea
\right),
\ee
where $\bZ=(\bA,\bPi,\pi)$
and 
$\bH_P$ denotes the reduced Hamiltonian
$\bH_P(\bA,\bPi,\pi):=\bH(\bA,\bPi,q,P,\pi)$.
According to (\ref{Ham2d}) and (\ref{Ham}),
\beqn\la{Ham22}
\bH_P(\bA,\bPi,\pi)
&=&\fr12\!\int\![|\bPi(y)|^2+|\curl \bA(y)|^2]dy
\nonumber\\
&&+\fr1{2m} [
P+\langle \bPi, \na_* \bA\rangle
\!-\!\langle \bA,\rho\rangle]^2
\!+\fr1{2I}[\pi-\langle y\we \bA(y),\rho(y)\rangle]^2.
\eeqn
In detail, the system (\ref{Hf3}) reads as
\be\la{Hf3d}
\left\{\ba{rcl}
\dot\bA(y,t)\!\!&\!\!=\!\!&\!\! D_\bPi \bH_P(\bZ(t))=\bPi(y,t)
+(v(t)\cdot\na) \bA(y,t)
\\
\dot\bPi(y,t)\!\!&\!\!=\!\!&\!\! -D_\bA \bH_P(\bZ(t))
=\De\bA(y,t)
+(v(t)\cdot\na) \bPi(y,t)+\bj(y,t)
\\
\dot\pi(t)\!\!&\!\!=\!\!&\!\! -\pi(t)\we D_\pi \bH_P(\bZ(t))=-\pi(t)\we\om(t)\ea\right|,
\ee
where 
$\bj(y,t)=j(q(t)+y,t)$
while
$v(t)$ and $\om(t)$ are given by 
 \be\la{PP3}
 v(t)=\fr1m
 [P-\langle \bPi(t) , \na_* \bA(t)\rangle-\langle\bA(y,t),\rho(y)\rangle],
 \qquad
  \om(t)=\fr1{I}[\pi(t)-\langle y\we \bA(y,t),\rho(y)\rangle].
 \ee
 The Hamiltonian (\ref{Ham22}) is
well-defined and Fr\'echet-differentiable on the reduced
phase space
\be\la{cZ}
 \cZ=\dot\cF^1\oplus \cF^0\oplus\R^3,
 \ee
 and it is conserved along trajectories of the system (\ref{Hf3d}).

\setcounter{equation}{0}
\section{The solitons}\la{sSol}
Here
we calculate the solitons 
 of the system (\ref{2ml2rd}), which 
are solutions of the form
\be\la{solvomA}
S_{v,\om}(t)=
(\bA_{v,\om}(x-vt),\bPi_{v,\om}(x-vt),vt, p_v,\pi_{v,\om}).
\ee
The second line of (\ref{2ml2rd}) and the definition (\ref{deo2}) imply that
\be\la{pqv}
p_v
=mv+\langle \bA_{v,\om}(y),\rho(y)\rangle,
\qquad
\pi_{v,\om}
=I\om+\langle y\we A_{v,\om}(q+y),\rho(y)\rangle.
\ee
The corresponding conserved (total) momentum is given by (\ref{mcoA}): 
\be\la{PJs}
P_{v,\om}=-\langle \bPi_{v,\om}, \na_* \bA_{v,\om}(t)\rangle+ mv+\langle\rho(y), \bA_{v,\om}(y)\rangle.
\ee
In the comoving frame,
the solitons  
 $ \bS_{v,\om}=(\bA_{v,\om},\bPi_{v,\om},\pi_{v,\om})\in\cZ$
  become stationary solutions
of the reduced system
(\ref{Hf3d}):
    \be\la{Hf3ds}
\left\{\ba{rcl}
0\!\!&\!\!=\!\!&\!\! D_\bPi \bH_{P_{v,\om}}(\bS_{v,\om})=\bPi_{v,\om}(y)
+(v\cdot\na) \bA_{v,\om}(y)
\\
0\!\!&\!\!=\!\!&\!\! -D_\bA \bH_{P_{v,\om}}(\bS_{v,\om})
=\De\bA_{v,\om}(y)+(v\cdot\na) \bPi_{v,\om}(y)+
\bj(y,t)
\\
0\!\!&\!\!=\!\!&\!\! -\pi_{v,\om}\we D_\pi \bH_{P_{v,\om}}(\bS_{v,\om})=-\pi_{v,\om}\we\om
\ea\right|.
\ee
The first two equations imply that
\be\la{solit2}
\De  \bA_{v,\om}(y)-(v\cdot\na)^2 \bA_{v,\om}(y)
=-\bj(y,t)=
-\cP[v\rho(\cdot)]
+\om \we y\rho(y)].
\ee
Note that $\cP[y\we\om\rho(y)]=y\we\om\rho(y)$ due to the spherical symmetry \eqref{rosym},
while on the Fourier transform side one has
$\widehat{\cP[v\rho]}(k)=(v-\fr{k(v\cdot k)}{k^2})\hat \rho(k)$.
Hence, 
solving \eqref{solit2} on the Fourier transform side, we obtain:
\be\la{solit3}
\hat  \bA_{v,\om}(k)=\fr{(v-\fr{k(v\cdot k)}{k^2}-i\om \we \na)\hat \rho(k)}{k^2-(v\cdot k)^2}.
\ee
Now
the first equation of  (\ref{Hf3ds}) gives 
$
\bPi_{v,\om}(y)=-v\cdot\na \bA_{v,\om}(y).
$
The conditions
(\ref{rosym}) imply that
\be\la{solH}
\bA_{v,\om}\in \dot\cF^1,\quad \bPi_{v,\om}\in \cF^0,\qquad {\rm for}\,\,\,|v|<1,\,\,\om\in\R^3.
\ee
 \br
 \rm
 For $|v|\ge 1$,
  formula (\ref{solit3}) shows that $\bA_{v,\om}\not\in \dot \cF^1$,
 hence the corresponding solitons $\bS_{v,\om}\in \cZ$ do not exist.
  
 \er
It remains to check the last equation in (\ref{2ml2rd}). It is equivalent to the relation
\be\la{piom2}
\pi_{v,\om}\Vert \om.
\ee
\bl\la{lmam}
Let all conditions (\ref{rosym}) hold.
Then the relation (\ref{piom2}) holds 
if and only if $(v,\om)\in\Si$. In this case,
\be\la{Ieff}
  \pi_{v,\om}=I_\eff\om,
\qquad I_\eff=
 I+\de I,\qquad\de I=
 \left\{\ba{ll} \de I^\Vert=
 \ds \int  \fr{k_3^2+k_2^2}
{[k^2-v^2 k_1^2]k^2}|\na\rho(k)|^2dk,&\om\Vert v
\\\\
\de I^\bot=
\ds \int  \fr{k_3^2+k_2^2}
{[k^2-v^2 k_2^2]k^2}|\na\rho(k)|^2 dk, &\om\bot v
\ea\right|.
\ee 
\el

\noindent
Above, $ I_\eff=I_\eff(v)$ is the  effective moment of inertia of the soliton.
We will prove this lemma in Appendix \ref{aA}.
By (\ref{rosym}), we have
\be\la{IIe}
I_\eff>I.
\ee

\bc
The soliton $\bS_{v,\om}\in\cZ$ exists and is unique 
if $(v,\om)\in\Si$. The solitons do not exist for $(v,\om)\not\in\Si$.
\ec
{
The following lemma is a particular case of  Lemma 5.2 from \ci{KK2023_LH}.

\bl\la{lPv}
The total  momentum (\ref{PJs}) 
admits the representation
\be\la{Pm}
P_{v,\om}=m_\eff(|v|,|\om|)v,\qquad (v,\om)\in\Si,
\ee
where the effective mass $m_\eff$ is a srictly increasing continuous
function of $|v|\in[0,1)$ and $|\om|\in [0,\infty)$.
\el

}

Now we formulate the definition of the orbital stability 
of solitons (\ref{solvomA}) 
in  terms of  reduced
trajectories
corresponding to solutions
$
Y(t)
=(A(x,t),\Pi(x,t),q(t),p(t),\pi(t))\in C(\R,\cY)
$
 of the system (\ref{2ml2rd}).
Introduce the distance
\be\la{dist}
 d( Y(t),S_{v,\om}(t)):=\Vert \bZ(t)- \bS_{v,\om}\Vert_{\cZ}
 +|\dot q(t)-v(Y(t))|+|\om-\om(Y(t))|.
\ee
Here
$
 \bZ(t)
 =( \bA(y,t), \bPi(y,t),\pi(t)) \in C(\R,\cZ),
$
where
 $\bA$, $\bPi$  are defined according to  (\ref{fcom}), while 
$v(Y(t))$, $\om(Y(t))$ are expressed by (\ref{PP3}). 
Suppose that
the initial data $Y(0)$ is close to $S_{v,\om}(0)$ in the following sense:
\be\la{tYt}
 d(Y(0),S_{v,\om}(0))<\ve.
\ee

\bd\la{dsta}
The  soliton (\ref{solvomA})
is orbitally stable if
for any $r>0$ there 
is $\ve>0$ such that, for any solution 
$Y(t)$
 of the system {\rm  (\ref{2ml2rd})}
the inequality {\rm (\ref{tYt})}
 implies that
\be\la{tYt2}
 d( Y(t), S_{v,\om}(t))<r,\qquad t\in\R.
\ee
\ed

Obviously, for $\rho(x)\equiv 0$ all solitons are orbitally stable; see
Remark \ref{r0}.


\setcounter{equation}{0}
\section{The Lyapunov function}
To prove the stability of a soliton $\bS_{v,\om}$, we must construct  a Lyapunov 
function $\Lam(\bZ)$ which is an invariant and admits a lower bound:
\be\la{lo-bound}
\de\Lam:=\Lam(\bS_{v,\om}+\de \bZ)-\Lam(\bS_{v,\om})\ge \vka\Vert \de\bZ\Vert_{\cZ}^2,\qquad \de\bZ\in \cZ,\qquad 
\Vert\de \bZ\Vert_\cZ\ll 1
\ee
with some $\vka=\vka(v)>0$.
We will construct the Lyapunov function as a perturbation
of the Hamiltonian $\bH_{v,\om}:=\bH_{P_{v,\om}}$ by a suitable Casimir 
invariant.

 In this section we calculate 
$\de\bH_{v,\om}=\bH_{v,\om}(\bS_{v,\om}+\de \bZ)-\bH_{v,\om}(\bS_{v,\om})$
for $\de\bZ=(\al(y),\beta(y),\ga)\in\cZ$, 
where
\be\la{dvab}
\dv\,\al(y)=\dv \,\beta(y)=0
\ee
according to  (\ref{AA}).
 We have 
\beqn \la{dH}
\de \bH_{v,\om}&=&
\fr12\int[|\bPi_{v,\om}+\beta|^2+|\curl (\bA_{v,\om}+\al)|^2]\, dy
-\fr12\int[|\bPi_{v,\om}|^2+|\curl \bA_{v,\om}|^2]\, dy
\nonumber\\
&&+\fr1{2m} [P_{v,\om}+\langle \bPi_{v,\om}+\beta, \na_*  
(\bA_{v,\om}+\al)\rangle-\langle \bA_{v,\om}(y)+\al,\rho(y)\rangle]^2
\nonumber\\
&&
-\fr1{2m} [P_{v,\om}+\langle \bPi_{v,\om}, \na_*  \bA_{v,\om}\rangle-\langle \bA_{v,\om}(y),\rho(y)\rangle]^2
\nonumber\\
&&
+\fr1{2I}[\pi_{v,\om}+\ga-\langle y\we (\bA_{v,\om}(y)+\al),\rho(y)\rangle]^2
-\fr1{2I}[\pi_{v,\om}-\langle y\we \bA_{v,\om}(y),\rho(y)\rangle]^2.
\nonumber
\eeqn
After rearrangements, we obtain 
\beqn \la{dH2}
&&\de \bH_{v,\om}=
\fr12\int(|\beta|^2+|\curl \al|^2)\, dy
+
\int (\bPi_{v,\om}\cdot \beta+\curl\bA_{v,\om}\cdot\curl\al)\,dy
\nonumber\\
&&
+\fr1{2m}[ (p_{v,\om}+\de p)^2
-p_{v,\om}^2]
+\fr1{2I}[(M_{v,\om}+\de M)^2
-M_{v,\om}^2],
\eeqn
where
\be\la{pvb}
\left\{\ba{rcl}
p_{v,\om}&:=&P_{v,\om}+\langle \bPi_{v,\om}, \na_*  \bA_{v,\om}\rangle-\langle \bA_{v,\om}(y),\rho(y)\rangle=mv
\\
\de p&:=&
\langle \beta, \na_*  \bA_{v,\om}\rangle
+\langle \bPi_{v,\om}, \na_*  \al\rangle
+\langle \beta, \na_*  \al\rangle-\langle \al(y),\rho(y)\rangle
\\
M_{v,\om}&:=&\pi_{v,\om}-\langle y\we \bA_{v,\om}(y),\rho(y)\rangle=I\om
\\
\de M&:=&\ga-\langle y\we \al,\rho(y)\rangle
\ea\right|
\ee
due to (\ref{PJs}) and (\ref{pqv}).
Using the first two equations of (\ref{Hf3ds}) and taking into account (\ref{dvab}), we obtain:
\beqn\la{rin2}
&&\int (\bPi_{v,\om}\cdot \beta+\curl\bA_{v,\om}\cdot\curl\al)\,dy=
\langle\bPi_{v,\om}, \beta\rangle-\langle  \De\bA_{v,\om},\al\rangle
\nonumber\\
&&=
-\langle  (v\cdot \na)\bA_{v,\om}, \beta \rangle-\langle  \bPi_{v,\om}, (v\cdot \na)\al \rangle
+\langle v\rho(y)+y\we\om \rho(y),\al\rangle
\nonumber\\
&&=-v\cdot\de p-\om\cdot\de M+\om\cdot\ga+
\langle\beta, (v\cdot \na)\al\rangle. \nonumber
\eeqn
Substituting into (\ref{dH2}), after rearrangements,  we get 
$
\de \bH_{v,\om}=\bJ_1+\bJ_2,
$
where
\beqn
\la{J1}
\bJ_1&=&\fr12\int(|\beta|^2+|\na \al|^2)dy+\langle\beta,v\cdot\na\al\rangle,\\
\la{J2}
\bJ_2&=&
\fr m{2}[ (v+\de p/m)^2-v^2-2v\cdot\de p/m]
+\fr I{2}[(\om\!+\!\de M/I)^2-\om^2-2\om\de M/I]+\om\cdot\ga
\nonumber\\
&=&
\fr {(\de p)^2}{2m}+ \fr {(\de M)^2}{2I}=\fr {(\de p)^2}{2m}+
\fr 1{2I}[\ga-\langle y\we \al,\rho(y)\rangle]^2+\om\cdot\ga.\la{dH51}
\eeqn
As a result,
\be\la{dH22}
\de \bH_{v,\om}=
\fr12\!\int\!(|\beta|^2\!\!+\!|\na \al|^2)dy\!+\!\langle\beta,v\!\cdot\!\na\al\rangle\!+\!
\fr {(\de p)^2}{2m}\!+\!
\fr1{2I}[\ga\!-\!\langle y\we \al,\rho(y)\rangle]^2+\om\cdot\ga.
\ee
The first  term $\bJ_1$ admits a lower bound 
\be\la{dH4}
\bJ_1\ge \fr{1-|v|}2 \int(|\beta|^2+|\na \al|^2)\, dy.
\ee


\setcounter{equation}{0}
\section{Orbital stability of solitons with $\om=0$}

In the case $\om=0$,
 the solitons $\bS_{v,0}$ are critical points of the reduced Hamiltonian  
  $\bH_{v,0}$:
\be\la{Lyap3}
D_\bA\bH_{v,0}(\bS_{v,0})=0,
\qquad D_\bPi\bH_{v,0}(\bS_{v,0})=0,
\qquad D_\pi\bH_{v,0}(\bS_{v,0})=0. 
\ee
Indeed,
the first and second identities hold due to the first
two equations of (\ref{Hf3ds}). 
The last identity
follows from
 (\ref{Ham22}) and (\ref{pqv}):
\be\la{Dom}
D_\pi \bH_{v,\om}(\bS_{v,\om})=
\fr1I [\pi_{v,\om}-\langle y\we \bA_{v,\om}(y),\rho(y)\rangle]
=\om.
\ee
So, in the case $\om=0$ we can consider the
Hamiltonian as the  Lyapunov function. Let us prove the lower bound (\ref{lo-bound}) for $\Lam=\bH_{v,0}$.
\bl \la{pb}
Let the conditions {\rm (\ref{rosym})}  hold and let $|v|<1$. Then
\be\la{lo-bound-H}
\de\bH_{v,0} :=\bH_{v,0}(\bS_{v,0}+\de \bZ)-\bH_{v,0}(\bS_{v,0})\ge \vka\Vert \de\bZ\Vert_{\cZ}^2,
\quad \de\bZ=(\al(y),\beta(y),\ga)\in \cZ,\quad \Vert\de \bZ\Vert_\cZ\ll 1
\ee
with some $\vka>0$.
 \el 
    \bpr
Due to  (\ref{J1})--(\ref{dH51}) with $\om=0$, it  suffices to check  that
\be\la{resu21}
\fr{1-|v|}2
\int |\na \al(y)|^2\,  dy
+
\fr1{2I}
 [\ga-\langle y\we \al(y),\rho(y)\rangle]^2
\ge 
\vka(\Vert \na\al\Vert_{L^2}^2+\ga^2).
\ee
Denote $\mu:=\ga-\langle y\we \al(y),\rho(y)\rangle$. 
Then 
(\ref{resu21}) can be rewritten equivalently  as
$$
\Vert \na\al\Vert_{L^2}^2+[\mu+\langle y\we \al(y),\rho(y)\rangle
]^2\le C(v,I)
(\int |\na \al(y)|^2\, dy+ \mu^2).
$$
It remains to note that the Cauchy--Schwarz inequality gives 
$$
|\langle   y\we \al(y),\rho(y)\rangle|=|\langle \na\hat\rho(k)\we \hat\al(k)\rangle| 
=|\langle \frac{\na\hat\rho(k)}{|k|}\we |k|\hat\al(k)\rangle| \le\Vert \na\al\Vert_{L^2} \Big[\int\frac{|\na\hat\rho(k)|^2}{k^2} dk\Big]^{1/2}. \qquad \quad
\qedhere
$$
\epr
 The lemma implies the orbital stability of solitons with $\om=0$:
\bt\la{t1}
Let  all the conditions {\rm (\ref{rosym})}  hold and let $|v|<1$.
Then 
the soliton $S_{v,0}$
is orbitally stable.
\et
\bpr
The proof requires a modification of the general scheme of \ci{BG1993}, \ci{IKM2004} and \ci{KS1998}.
Namely, 
the inequality (\ref{tYt}) with sufficiently small $\ve>0$
implies that the momentum $P$ of the solution
$Y(t)$ is close to $P_{v,0}$:
\be\la{mclo}
 |P-P_{v,0}|
\to 0,\qquad
 \ve\to 0.
\ee
{
It is important, that
 the map $v\mapsto P_{v,0}$ is a homeomorphism
  $\R^3\to\R^3$ by  Lemma \ref{lPv}.
 Hence,  $P=P_{v_*,0}$, where 
 \be\la{mclo2}
 |v_*-v|
\to 0,\qquad
 \ve\to 0.
 \ee
Combining (\ref{mclo}) with  (\ref{tYt}), we obtain 
$
\Vert \bZ(0)-\bS_P\Vert_{\cZ}
\to 0$ as $\ve\to 0.
$
Therefore, 
$$
0\le H_P(\bZ(0))-H_P(\bS_P)\le \ve_1,
\qquad \ve_1\to 0\,\,\,{\rm as}\,\,\,\ve\to 0.
$$
Hence,
the 
conservation of  $H_P(\bZ(t))$
for the system (\ref{Hf3d}) with the parameter $P$
 implies that
$$
0\le H_P(\bZ(t))-H_P(\bS_P)\le \ve_1,\quad t\in\R.
$$
Now we apply the lower bound {\rm (\ref{lo-bound-H})} with $v=v_*$:
\be\la{lo-bound2}
\bH_P(\bS_P+\de \bZ)-\bH_P(\bS_P)\ge \vka
\Vert \de\bZ\Vert_{\cZ}^2,\qquad 
\de\bZ\in \cZ,\qquad 
\Vert\de \bZ\Vert_\cZ\ll 1.
\ee
Setting $\de\bZ=\bZ(0)-\bS_P$,
we obtain
$
\sup_{t\in\R}
\Vert \bZ(t)-\bS_P\Vert_{\cZ}
\to 0$ as $\ve\to 0.
$
Finally, combining with  (\ref{mclo2}), we get
}
\be\la{crp6}
\sup_{t\in\R}
\Vert \bZ(t)-\bS_{v,0}\Vert_{\cZ}
\to 0,\qquad\ve\to 0.
\ee
It remains to notice that
the last terms on the right-hand side of (\ref{dist}) 
are small uniformly in time:
\be\la{ter}
\sup_{t\in\R}[|\dot q(t)-v|+|\om(t)|]\to 0,\qquad\ve\to 0.
\ee
This follows from (\ref{PP3})  together with
(\ref{mclo}) and (\ref{crp6}).
\epr
\setcounter{equation}{0}
\section{Angular \!momentum \!conservation \!for \!reduced \!system \!with \!$P=0$}
\la{som}

 For $\om\ne 0$, the solitons $\bS_{v,\om}$ are not critical points of the reduced Hamiltonian
because of
(\ref{Dom}). In this case
we will consider the Lyapunov function which is a perturbation
of   the Hamiltonian $\bH_{v,\om}$ by a  Casimir invariant:
\be\la{Lyap2*}
\Lam_{v,\om}(\bZ)=\bH_{v,\om}(\bZ)-\fr{\nu_{\rm eff}}2{\pi^2},
\qquad \nu_{\rm eff}:=\fr1{I_{\rm eff}}.
\ee
 This function  is an invariant for the reduced system (\ref{Hf3d}),
and the solitons $\bS_{v,\om}$ are its critical points since
$
 {D_\pi \Lam(\bS_{v,\om})=\om-\nu_{\rm eff}I_{\rm eff}\om=0}
$
by (\ref{Dom}) and  (\ref{Ieff}).
Now 
(\ref{dH22}) leads to
\beqn\la{deL}
\de\Lam_{v,\om}(\de\bZ)\!\!&\!\!\!:=\!\!\!&\!\!\Lam(\bS_{v,\om}+\de \bZ)-\Lam(\bS_{v,\om})=\de\bH_{v,\om}
-\fr{\nu_\eff}2((\pi_{v,\om}+\ga)^2-\pi_{v,\om}^2)
\nonumber\\
\!\!&\!\!\!=\!\!\!&\!\!\fr12\!\int\!(|\beta|^2\!\!+\!|\na \al|^2)dy\!+\!\langle\beta,v\!\cdot\!\na\al\rangle\!+\!
\fr {(\de p)^2}{2m}\!+\!
\fr \nu 2[\ga\!-\!\langle y\we \al,\rho(y)\rangle]^2\!-\!\fr{\nu_\eff}2\ga^2,\quad \nu\!:=\!\fr1I.
\qquad
\eeqn
The presence of the last  ``negative" term makes problematic
a lower bound of type  (\ref{lo-bound}),
 and
perturbations of $\bH_{v,\om}$
by other  Casimir invariants  do not help; see Appendix \ref{sLJ}.

That is why we need to involve an 
additional invariant of our system: the angular momentum.
 However, the expression (\ref{amcoA}) is a functional on the phase space $\cZ$
of the reduced system only in the case of zero total momentum $P=0$.
This is why we consider 
below
solutions and solitons 
of the system  (\ref{2ml2rd})
 with the total momentum $P=0$.
In this case,
 the reduced system (\ref{Hf3d}) reads
\be\la{2ml2rd*}
\dot \bA=\bPi\!+\!(v\cdot\na) \bA,
\qquad
\dot \bPi=\De \bA+(v\cdot\na) \bPi
+\bj,
\qquad
 \dot \pi= -\pi\we\om,
\ee
where 
 $v=v(t)$ and $\om=\om(t)$ are given by (\ref{PP3}) with $P=0$.
The angular momentum conservation (\ref{amcoA}) with $P=0$  becomes
\be\la{jcon}
 J(\bZ(t)):=\langle \bA(y,t)\we\bPi(y,t)\rangle-
\langle (y\we \na)_*\bA(y,t),\bPi(y,t)\rangle+\pi(t)=\const,
\quad t\in\R.
\ee
 Stationary solutions of this system are the solitons $\bS_{v,\om}$ with  $v=0$.
 This follows from the formula $P_{v,\om}=m_\eff(v,\om)v$
established in  \ci[(5.17)]{KK2023_LH}. 
 We denote the solutions as $\bS_\om:=\bS_{0,\om}$.
 Due to  the first equation of (\ref{Hf3ds}) with $v=0$, 
\be\la{Som}
\bS_\om=(\bA_\om, 0,\pi_\om)=(\bA_{0,\om}, 0,\pi_{0,\om}).
\ee 
Finally, the Hamiltonian 
(\ref{Ham22}) with $P=0$
becomes
 \beqn\nonumber
\bH_0(\bZ)&=&\fr12\!\int[|\bPi(y)|^2+|\curl \bA(y)|^2]dy\\
\la{Ham*}
&&+
\fr1{2m} [\langle \bPi, \na_* \bA\rangle-\langle \bA,\rho\rangle]^2
+\fr1{2I}[\pi-\langle y\we \bA(y),\rho(y)\rangle]^2, \quad \bZ=(\bA,\bPi,\pi).
\eeqn

  The set  $\cZ_0^\infty=\cZ\cap [C_0^\infty\oplus C_0^\infty\oplus\R^3]$ is dense in $\cZ$ by Remark \ref{rde}.
    Proposition \ref{pwpA} iii) implies the following lemma.
   \bl\la{lemJ}
   For solutions to (\ref{2ml2rd*}) with 
   initial state $(\bA(y,0),\bPi(y,0),\pi(0))\in\cZ_0^\infty$, we have
   for any $T>0$, $l=0,1,2$ and
$|\al|+l\le 2$, 
    that
   $\pa_y^\al \pa_t^l  \bA(x,t)\in C^{2-(|\al|+l)}(\R^3\times[0,T])\!\otimes\!\R^3$, and 
\be\la{AP1-new}
\left\{\ba{rcll}
|\pa_t^l \bA(y,t)|&\le& C(1+|y|)^{-2}, &l\le2
\\\\
|\pa_y^\al\pa_t^l  \bA(y,t)|&\le& C(1+|y|)^{-3},&\al\ne 0,\,\,|\al|+l\le 2
\ea\right|,\qquad y\in\R^3,\,\,\, |t|\le T,
\ee
      and the angular momentum conservation (\ref{jcon}) holds.
   \el

The conservation (\ref{jcon}) suggests
that
the stability of solitons $\bS_\om$ for the nonlinear system (\ref{2ml2rd*})
 follows from a lower bound  
of type (\ref{lo-bound})
 on the manifold
 \be\la{M}
\cM_\om:=\{\bZ\in \cZ:J(\bZ)=J(\bS_\om)\}
\ee
for the function (\ref{deL}) with $v=0$.
However, the expression (\ref{jcon}) is 
  not well-defined on the phase space $\cZ$, so 
the manifold $\cM_\om$ is also not well-defined.
On the other hand,
the ``tangent space" $\aT_\om=T_{\bS_\om}\cM_\om$ can be  defined as
\be\la{TM}
\aT_\om:=\{\de\bZ\in \cZ: (E_{\om,n}, \de\bZ)=0,\,\,\,n=1,2,3\},
\qquad E_{\om,n}:=dJ_n(\bS_\om)=(0,b_{\om,n}(x),1)\in\cZ^0,
 \ee
 where $\cZ^0:=\cF^0\oplus \cF^0\oplus\R^3$. Indeed,
according to
 (\ref{jcon}), 
 \beqn\la{dJ}
 (dJ_n(\bS_\om),\de \bZ)&=&
 \langle (\bA_\om(y)\we\beta(y))_n\rangle-
\langle (y\we \na)_n\bA_\om(y),\beta(y)\rangle
+\ga_n
\nonumber\\
&=&\langle(\hat\bA_\om(k)\we\hat\beta(k)_n\rangle+
\langle (\na\we k)_n\hat\bA_\om(k),\hat\beta(k)\rangle
+\ga_n,\quad \de\bZ\!=\!(\al,\beta,\ga)\!\in\!\cZ,
\qquad\eeqn
 where $\bA_\om(y)$ and $(y\we \na)_n\bA_\om(y)$ belong to $L^2$
 by (\ref{solit3}) with $v=0$. Moreover, 
 \be\la{En}
E_{\om,n}=(0,b_{\om,n}(x),1)\in \cZ^0\cap [H^\infty\oplus H^\infty\oplus\R^3].
\ee
Indeed,  (\ref{rosym}) and
 (\ref{solit3}) with $v=0$ imply
that 
\be\la{Eb}
(1+|k|)^s\hat  \bA_{\om}(k)\in L^2,\quad 
(1+|k|)^s(\na\we k)_n\hat \bA_\om(k)\in L^2,
\qquad \forall s>0.
\ee

\bd
Denote by $\aT_\om^0$ the Hilbert space which is the closure of $\aT_\om\cap \cZ^0$ in $\cZ^0$ endowed with the norm of $\cZ^0$.
\ed
 
 The space $\aT_\om^0$ admits also the representation (\ref{TM}) with $\de\bZ\in \cZ^0$.
So,
$\aT_\om^0$
  is a closed linear subspace of codimension three in $\cZ^0$.
 Denote by $P_\om$ the orthogonal 
 projection of $\cZ^0$ onto $\aT_\om^0$: 
\be\la{Pom}
P_\om=1-\sum_1^3 p_nE_{\om,n}\otimes E_{\om,n}^*,
\ee
where $p_n$ are suitable normalization factors.


\setcounter{equation}{0}
\section{The stability of the linearized dynamics in the tangent space}
Let us linearize the reduced system
(\ref{2ml2rd*})  
 at the soliton $\bS_\om$. 
 We write the system  as
 \be\la{2ml2rd*F}
 \dot \bZ(t)=F(\bZ(t)),\qquad t\in\R;\qquad \bZ(t)=(\bA(t),\bPi(t),\ga(t)).
 \ee
The {\it variational equation} is obtained 
 by replacing 
$\bZ(t)$ with $\bZ(t)+\xi(t)$
and
keeping the terms linear in $\xi(t)$. Formally, the variational equation reads as
 $
 \dot\bxi(t)=dF(\bZ(t))\bxi(t).
 $
 Accordingly, 
 the linearization of  (\ref{2ml2rd*F}) at the soliton is
 \be\la{leq}
 \dot\bxi(t)=dF(\bS_\om)\bxi(t).
 \ee
Let us calculate the  corresponding Hamilton--Poisson structure for the linearized
dynamics.
 Substitute $\bZ(t)=\bS_\om+\xi(t)$
 into (\ref{Hf3}). Then (\ref{Hf3}) implies that
 \be\la{Hf32}
 \dot\xi(t)=
 \cbJ(\bS_\om)DH_0(\bS_\om+\xi(t))+\cbJ_0(\bS_\om+\xi(t)) DH_0(\bS_\om+\xi(t)),\quad \cbJ_0(\xi(t))\!=\!\!\left(\!\!
\ba{ccccc}
0&0&0\\
0&0&0\\
0&0&-\ga(t)\we
\ea
\!\!\right)\!.
 \ee
 We have to keep in the right-hand side of (\ref{Hf32})
 only linear terms in $\xi$. 
 Thus,
 it suffices to keep in  $H_0(\bS_\om+\xi(t))$ only linear and quadratic terms.
 Hence,
 we can replace $H_0(\bS_\om+\xi)$ with $\xi=(\al,\beta,\ga)\in\cZ$ by
 its quadratic part: according to (\ref{dH22}) and (\ref{pvb}),
 \be\la{repH}
 \ti H_\om(\xi)=\fr12\!\int\!(|\beta|^2\!\!+\!|\na \al|^2)dy\!+\fr1{2m} {(\langle \beta, \na_*  \bA_{0,\om}\rangle
\!-\!\langle \al(y),\rho(y)\rangle)^2}\!+\!
\fr \nu 2[\ga\!-\!\langle y\we \al,\rho(y)\rangle]^2+\om\cdot\ga.
 \ee
Note that the last linear term contributes only to the last component
of the differential, 
 $
 D_\ga \ti H_0(\xi)=\nu [\ga\!-\!\langle y\we \al,\rho(y)\rangle]+\om.
  $
  Moreover, due to the structures of the matrices $\cbJ(\bS_\om)$ and 
  $\cbJ_0(\xi)$,
this component contributes only to the equation for $\dot \ga(t)$:
\beqn\la{dog}
\dot\ga(t)&=&-\pi_\om\we(\nu [\ga\!-\!\langle y\we \al,\rho(y)\rangle]+\om)
-\ga\we(\nu [\ga\!-\!\langle y\we \al,\rho(y)\rangle]+\om)
    \nonumber\\
    &=&-\nu\pi_\om\we [\ga\!-\!\langle y\we \al,\rho(y)\rangle]
-\ga\we(\nu [\!-\!\langle y\we \al,\rho(y)\rangle]+\om)
    \eeqn
    since $\pi_\om\we\om=0$ by (\ref{Ieff}). 
    The corresponding  linearized equation is
\be\la{dog2}
\dot\bga(t)
    =-\nu\pi_\om\we [\bga\!-\!\langle y\we \bal,\rho(y)\rangle]
-\bga\we\om.
    \ee
    Now it is clear that
    the linearized equation (\ref{leq}) reads as
    \be\la{Hf33}
 \dot\bxi(t)=
 \cbJ(\bS_\om)D\ti H_0(\bxi(t))-\left(\!\!\!\ba{c}0\\0\\\bga\we\om\ea\!\!\!\right),\qquad \bxi(t)=(\bal(t),\bbeta(t),\bga(t))\in\cZ.
  \ee
Now we are going to prove that the tangent space $\aT_\om$  is invariant with respect 
 to the linearized dynamics (\ref{leq}). As the first step, we prove
that the vector field $dF(\bS_\om)\bxi$ is tangent to $\aT_\om$.
 \bl\la{linvar}
 For $\bxi\in \cZ$, one has
 \be\la{dJF2}
  (dJ_n(\bS_\om),dF(\bS_\om)\bxi)=
  0,\qquad n=1,2,3.
  \ee  
  \el
  \bpr
    Let us consider solutions $\bZ(t)$ to (\ref{2ml2rd*F}) 
  with initial states $\bZ(0)\in\cZ_0^\infty$. Then 
   the angular momentum conservation  (\ref{jcon}) holds by Lemma \ref{lemJ}.
    Differentiating  (\ref{jcon}) in time, we obtain by (\ref{En}) that
 $
 ( dJ_n(\bZ(t)),\dot\bZ(t))=(dJ_n(\bZ(t)),F(\bZ(t)))=0$ for $t\in\R.
 $
 Therefore, 
 \be\la{dJF3}
 (dJ_n(\bZ),F(\bZ))=0,\qquad \bZ\in \cZ_0.
 \ee
 Taking into account (\ref{En}) and the
 density of $\cZ_0^\infty$ in $\cZ$, 
 we 
 conclude that  the identity (\ref{dJF3})
 holds 
  for 
 $\bZ=\bS_\om+\ve\bxi$ with any vector
  $\bxi\in \cZ$. 
  Differentiating 
  the obtained identity 
  in $\ve$ at $\ve=0$,
  we get
 \be\la{dJF4}
  ( [L_{\bxi} dJ_n](\bS_\om),F(\bS_\om))+ (dJ_n(\bS_\om),[L_{\bxi} F](\bS_\om))=0,
   \ee
   where $L_{\bxi}$ stands for the  operator of
differentiation.
   However,      $F(\bS_\om)=0$, so
 (\ref{dJF2}) follows  since
  $dF (\bS_\om)\bxi:= [L_{\bxi} F](\bS_\om)$.
  \epr

 The following theorem on stability is the main result of the present paper.
 
 \bt\la{tlos}
 Let conditions (\ref{rosym}) 
hold. Then
\smallskip\\
 i)
For every $\bxi(0)\in\aT_\om$ there exists a unique solution 
$\bxi(t)\in C(\R,\aT_\om)$ to (\ref{Hf33}), and the map
$\bxi(0)\mapsto \bxi(t)$ is continuous in $\aT_\om$
for $t\in\R$.
ii) There is the uniform bound
\be\la{aprlin}
\Vert\xi(t)\Vert_\cZ\le C\Vert\xi(0)\Vert_\cZ,\qquad t\in\R.
\ee

\et
\bpr
It suffices to prove (\ref{aprlin}) as an priori estimate;
then
 the theorem will follow by the same methods as Proposition \ref{pwpA}.
 The estimate holds due to the conservation of an appropriate ``energy functional". Let us show that the  functional 
 can be chosen as
 the quadratic part of the Lyapunov function (\ref{Lyap2*}) with $v=0$:
 \be\la{Lyap2*a}
L_\om(\bxi)=\ti H_0(\bxi)-\fr{\nu_{\rm eff}}2{\bga^2}.
\ee
Indeed,
 (\ref{Q}) shows that the needed lower bound holds.
 It suffices to prove that the functional is conserved
 under the linearized dynamics (\ref{Hf33}).
 Assuming ``a priori" the existence of the solution and 
 differentiating $L_\om(t)=L_\om(\bxi(t))$, we obtain by (\ref{Hf33}) and (\ref{dog2}) that
   \beqn\la{dQ}
\dot L_\om(t)&=&(DL_\om(\bxi(t)),\dot\bxi(t))
\nonumber\\
&=&
\Big(D \ti H_0(\bxi(t))-
\left(\!\!\!\ba{c}0\\0\\ \nu_\eff\bga \ea\!\!\!\right),
\cbJ(\bS_\om)D\ti H_0(\bxi(t))\Big)
-\Big(DL_\om(\bxi(t)),\left(\!\!\!\ba{c}0\\0\\\bga\we\om\ea\!\!\!\right)\Big)
\nonumber\\
&=& -\nu_\eff\bga \cdot [-\pi_\om\we D_\ga\ti H_0(\bxi(t))]-[D_\ga
\ti H_0(\bxi(t))-\nu_\eff\ga]
\cdot(\bga\we\om)=0.\quad\quad \qedhere
\eeqn
Finally, 
the conservation of $L_\om$ provides (\ref{aprlin}) by
the following proposition.

\bp\la{plos}
The Lyapunov function $L_\om$ admits a lower bound: for some $\vka>0$,
\be\la{Qp}
L_\om(\bxi)
 \ge \vka\Vert\bxi\Vert_\cZ^2,
 \qquad \bxi\in \aT_\om.
\ee

\ep

 This theorem is proved in the remaining part of the paper.
    \epr

   \br\la{rec}
   \rm
   The calculation (\ref{dQ}) shows that the ``linearized" Lyapunov function (\ref{Lyap2*a})
   is conserved, but the terms $\ti H_\om(\bxi)$ and $\fr{\nu_{\rm eff}}2{\bga^2}$
   do not. This fact is contrary to the case of the ``nonlinear" Lyapunov function
   (\ref{Lyap2*}), where both terms are conserved quantities. This fact is a 
   peculiarity of the energy-Casimir method for the linearized system.
  
   \er


    \setcounter{equation}{0}
\section{The Kato--L\"owner--Heinz inequality }
In this section we prove
the lower bound (\ref{Qp}).
We will prove a stronger  bound than (\ref{Qp}):
\beqn\la{Q}
\cQ_\rho(\bxi) &:=&
\fr12\int[|\beta(y)|^2+|\na \al(y)|^2]dy
+
\fr \nu 2[\ga-\langle y\we \al,\rho(y)\rangle]^2-\fr{\nu_\eff}2\ga^2
\nonumber\\
&\ge& \vka\Big[
 \fr12
 \int [|\beta(y)|^2+|\na \al(y)|^2]dy+\ga^2\Big],\qquad \bxi=(\al,\beta,\ga)\in \aT_\om.
\eeqn
The quadratic form $\cQ_\rho$ is continuous
on the Hilbert space $\aT_\om$ endowed with the norm of $\cZ$. Hence,
 it suffices to prove the bound (\ref{Q}) on the space of smooth functions 
$\aT_\om^\infty:=\aT_\om\cap\cZ^\infty_0$
which is dense in $\aT_\om$.
For the smooth functions, the bound (\ref{Q}) can be written as
\be\la{Q3}
(\bxi,Q _\rho \bxi)\ge \vka( \bxi, Q_0\bxi),\qquad \bxi\in\aT_\om^\infty,
\ee
where
the brackets $(\cdot,\cdot)$ denote the inner product in the Hilbert space $\cZ^0$,
and
\be\la{Q2}
Q_\rho =\left(
\ba{ccc}
- \De+\nu|m^*(x)\rho(x)\rangle \langle\rho(y) m(y)|
&
0
&
-\nu|m^*(x)\rho(x)\rangle
\\
0&1&0
\\
-\nu\langle\rho(y) m(y)|&0&\de
\ea
\right),\quad
Q_0=\left(\!\!\!
\ba{ccc}
-\De&0&0
\\
0&1&0
\\
0&0&\de
\ea
\!\!\right).
\ee
Here
 $\de:=\nu-\nu_\eff>0$ by (\ref{IIe}), and 
$m(x)$ denotes the operator $\al\mapsto x\we \al$, 
$\al\in\R^3$, with
the corresponding  skew-symmetric matrix.
The operator 
$Q_0$ is positive-definite and  selfadjoint  in $\cZ^0$ with domain $\cD^0$.
Hence, 
$Q_\rho $ is also a selfadjoint operator in $\cZ^0$ with the same domain since the difference
$Q_\rho -Q_0$ is a bounded finite rank operator in $\cZ^0$
by  (\ref{Q2}) and
 (\ref{rosym}).
 The common domain of both operators $Q_\rho $ and $Q_0$ contains the space
$
\cZ_0^\infty=\cZ\cap [C_0^\infty\oplus C_0^\infty\oplus\R^3]\subset \cZ^0.
$
The space is dense in $\cZ^0$ by Remark \ref{rde}.
In the next section, we will prove the following  proposition.

\bp
\la{ppos} 
 Let conditions (\ref{rosym}) hold.
 Then 
 \smallskip\\
 i)
 the  operator $Q_\rho $ is nonnegative:
\be\la{Qnn}
 (\bxi, Q_\rho \bxi)\ge 0,\qquad \bxi\in
 { \cZ_0^\infty. } 
\ee
ii) { $\dim\ker Q_\rho =3$} and $\ker Q_\rho \cap\aT_\om^0=0$.

\ep
\noindent{\bf Proof of Proposition \ref{plos}.}
Consider the restrictions of the quadratic forms  $( \bxi, Q_\rho \bxi)$
and $( \bxi, Q_0\bxi)$ onto {  $\aT_\om^\infty:=\cZ_0^\infty\cap \aT_\om$} and
denote by $\ti Q_\rho $ and $\ti Q_0$ the corresponding nonnegative
selfadjoint operators in $\aT_\om^0$:
\be\la{tiQ}
( \bxi, \ti Q_\rho \bxi)=( \bxi, Q_\rho \bxi),\qquad ( \bxi, \ti Q_0\bxi)=(\bxi, Q_0\bxi), \qquad \bxi\in \aT_\om^\infty.
\ee
The space {   $ \aT_\om^\infty$  is dense in $\aT_\om^0$} and
the operators can be expressed as
\be\la{tiQ2}
\ti Q_\rho =P_\om Q_\rho,\qquad \ti Q_0=P_\om Q_0,
\ee
where $P_\om$ is the projection  (\ref{Pom}).
\bl
\la{linv}
Let conditions (\ref{rosym}) hold.
Then
 $\ti Q_\rho \gg \mu\ti Q_0$ with sufficiently small $\mu>0$, i.e.
\be\la{QQ2}
\Vert \ti Q_\rho \bxi\Vert_{\cZ^0}\ge \mu\Vert \ti Q_0\bxi\Vert_{\cZ^0},
\qquad \bxi\in \aT_\om^\infty.
\ee
\el
\bpr
We have $\ti  Q_0=\ti Q_0 \ti Q_\rho ^{-1}\ti Q_\rho $. Hence,  
it suffices to prove that
 the operator $\ti Q_0\ti Q_\rho ^{-1}: \aT_\om^0\to \aT_\om^0$ is bounded
 since then
 $\Vert\ti  Q_0 \bxi\Vert_{\cZ^0}\le C \Vert\ti  Q_\rho  \bxi\Vert_{\cZ^0}$
 for $\bxi\in\aT_\om^\infty$
 which is equivalent to  (\ref{QQ2}).

 The difference $Q_\rho-Q_0$ is a finite rank bounded operator in $\cZ^0$. Hence, 
$\ti Q_\rho-\ti Q_0$ is a finite rank bounded operator in 
$\aT_\om^0$ by (\ref{tiQ2}) and (\ref{Pom}) since $E_{\om,n}\in \cZ^0$ by (\ref{TM}). Therefore, the  selfadjoint operators 
$\ti Q_\rho$ and $\ti Q_0$ have common domain 
$\ti\cD^0$. Moreover, Proposition \ref{ppos} ii)
implies that $\ti Q_\rho$ is injective, and 
hence it is a bijection
of $\ti\cD^0$ onto $\aT_\om^0$. Therefore,
the operator $\ti Q_0\ti Q_\rho^{-1}$ is well-defined on the entire Hilbert
space  $\aT_\om^0$, so the operator is bounded by the closed graph theorem.
\epr

 By Proposition \ref{ppos},  $\ti Q_\rho $ is a selfadjoint nonnegative operator in 
  $\aT_\om^0$, similarly to $\ti Q_0$.
 Hence, (\ref{QQ2}) and
 the Kato--L\"owner--Heinz inequality \ci[Theorem 2]{K1952} imply that 
 $\ti Q_\rho ^s\gg \mu^s\ti Q^s_0$ 
 for all  $s\in [0,1]$, i.e.,
 \be\la{Qs}
\Vert \ti Q_\rho ^s\bxi\Vert_{\cZ^0}\ge \mu^s\Vert \ti Q_0^s\bxi\Vert_{\cZ^0},
\qquad \bxi\in \aT_\om^\infty.
\ee
Finally, 
 this bound with $s=1/2$ obviously implies (\ref{Q3}) with
 $\vka=\mu$. $\hfill\Box$

 \br\la{rH}
 \rm
 
 Proposition \ref{ppos} eliminates the point spectrum
at zero
  of the operator $\ti Q_\rho$ in the tangent space $\aT_\om^0$. 
  However, this fact alone
 does not imply the bound  (\ref{Q}) because of presence of 
 the continuous spectrum of $\ti Q_\rho$ 
 { which extends up to the point zero.
The application of the Kato--L\"owner--Heinz inequality allows us 
 to establish the equivalence 
  of continuous spectra of $\ti Q_\rho$ and  
  $\ti Q_0$.
 }
 
 \er


\setcounter{equation}{0}
\section{The positivity of the Lyapunov function on the tangent space}
Here we prove Proposition \ref{ppos}.
It suffices to prove the nonnegativity (\ref{Qnn}) for $\beta=0$:
$$
\cB(Y)\!:=\!\!
 \int\! |\na \al(y)|^2dy\!+\!
 \nu[\ga\!-\!\langle m(y)\al(y),\rho(y)\rangle]^2
\! - \!\nu_\eff\ga^2\ge 0,\quad Y\!= \!(\al,\ga)\!\in\!  
 \cY^\infty\!:=\! [\dot\cF^1\!\cap \!C_0^\infty]\oplus\R^3.
$$
This form can be written as
$
\cB(Y)=(Y, BY)$ for  $Y\in \cY^\infty,
$
where the brackets $(\cdot,\cdot)$ now denote the inner product
in the Hilbert space
$\cY^0:=\cF^0\oplus\R^3$ and 
 \be\la{B2}
B=\left(
\ba{cc}
- \De+\nu|m^*\rho(x)\rangle\langle\rho(y) m|&
-\nu|m^*\rho(x)\rangle
\\
-\nu\langle\rho m(y)|&\de
\ea
\right)
\ee
by (\ref{Q2}).
Proposition \ref{ppos} i) follows from the next lemma.
\bl
\la{lpos} 
 Let conditions (\ref{rosym})  hold.
 Then 
 the  operator $B$ is nonnegative:
 \be\la{YB}
 (Y, BY)\ge 0,\qquad Y\in \cY^\infty.
\ee
\el

\bpr
For $\al\in \dot\cF^1\cap C_0^\infty $, the action of
the Laplacian  $\De$  in the sense of distributions coincide with 
the Friedrichs closure
of $\De$ in the invariant space $\cF^0$.
Hence, $B$ in (\ref{YB}) can be considered as a
selfadjoint  operator in $\cY^0$ with domain $D(B)$.
Therefore,
it suffices to show that for $\lam < 0$ the resolvent $R(\lam)=(B-\lam)^{-1}:\cY^0\to\cY^0$ is a bounded operator.
The operator $B-\lam$ is  selfadjoint in $\cY^0$.
Hence,  it remains to check that
\be\la{QQ3}
\ker (B-\lam)=0,\qquad\lam<0.
\ee
The calculation of 
the kernel  reduces to a finite-dimensional 
problem in a standard way. Indeed,
the equation 
{ $(B-\lam)Y=0$
with $Y=(\al,\ga)\in D(B)$}  reads as
\be\la{QXF}
\left\{\ba{rcl}
(-\De-\lam)\al+|\nu m^*\rho\rangle\langle\rho m|\al\rangle
-\nu m^*\rho\ga&=&0
\\
-\nu\langle\rho m|\al\rangle+(\de-\lam)\ga&=&0
\ea\right|.
\ee
Using the last equation, we rewrite the first one as
\beqn
(- \De-\lam)\al(x)&=&-m^*(x)\rho(x)(\de-\lam)\ga+\nu m^*(x)\rho(x)\ga
=m^*(x)\rho(x)(\nu_\eff+\lam)\ga,\,\, \lam\in\R.\qquad\qquad \la{Res20}
\eeqn
Hence, on the Fourier transform side,
\be\la{Res2}
\hat\al(k)=(k^2-\lam)^{-1}[m^*(-i\na)\hat\rho(k)(\nu_\eff+\lam)\ga],\qquad \lam\le 0.
\ee
Let us emphasise that the formula holds for $\lam=0$ since 
$[m^*(-i\na)\hat\rho](0)=0$.

\br\rm
For any  $w\in\R^3$, the vector field $m^*(x)\rho(x) w$ is divergence-free
by the spherical symmetry (\ref{rosym}). 
\er

Substituting (\ref{Res2}) into the second  equation of (\ref{QXF}),
we obtain:
\be\la{Res4-1}
A(\lam)\ga=0,
\quad
A_{j,l}(\lam):=-\nu\langle m(-i\na_j)\hat\rho(k) ,(k^2-\lam)^{-1}[m^*(-i\na_l)\hat\rho(k)]\rangle(\nu_\eff+\lam)+(\de-\lam).
\ee
The matrix $A(\lam)$ is Hermitian for $\lam\le 0$, so it is defined uniquely by its quadratic form: 
\beqn\la{Aqf}
\ga\cdot(A(\lam)\ga)
&=&-\nu\langle \na\hat\rho(k)\we\ga,(  k^2-\lam)^{-1}\na\hat\rho(k)\we\ga\rangle(\nu_\eff+\lam)+(\de-\lam)\ga^2
\nonumber\\
&=&-\nu\ga^2
\int
\fr{|\na\hat\rho(k)|^2 \sin^2\widehat{(k,\ga)}}{  k^2-\lam}dk
(\nu_\eff+\lam)+(\de-\lam)\ga^2
=-a_-(\lam)\ga^2+a_+(\lam)\ga^2.\nonumber
\eeqn
Thus, the matrix $A(\lam)$ is the scalar $-a_-(\lam)+a_+(\lam)$, and (\ref{QQ3})
is equivalent to 
\be\la{al}
a_-(\lam)\ne a_+(\lam),\qquad \lam< 0.
\ee
We note that the function $a_-(\lam)$ is  strictly increasing
while $a_+(\lam)$ is strictly decreasing for $\lam\le 0$, and
that both functions are continuous. Hence,  (\ref{al})
 holds since
$a_-(0)= a_+(0)$. Indeed, the last equality is equivalent to
\be\la{al3}
\nu\nu_\eff
\int
\fr{|\na\hat\rho(k)|^2 \sin^2\widehat{(k,\ga)}}{  k^2}dk
= \de=\nu-\nu_\eff=\nu\nu_\eff\de I,
\ee
which
 holds since
$\ds\int
\fr{|\na\hat\rho(k)|^2 \sin^2\widehat{(k,\ga)}}{  k^2}dk
= \de I$
by (\ref{Ieff}) with $v=0$.
    \epr

 \noindent{\bf Proof of Proposition \ref{ppos}.}
 The nonnegativity (\ref{Qnn}) follows from Lemma \ref{lpos}.
 By (\ref{Q2}), $\ker Q_\rho=\{(\al_\ga,0,\ga)\}$, where $\ga\in\R^3$, and
 $\al_\ga$ is defined by (\ref{Res2}) with $\lam=0$:
 \be\la{Res202}
 \hat\al_\ga(k)=\nu_\eff|k|^{-2}m^*(-i\na)\hat\rho(k)\ga,
 \ee
  It is
 important that $\al_\ga\in\dot\cF^1$ since 
$[m^*(-i\na)\hat\rho](0)=0$.
Hence,  $\dim \ker Q_\rho=3$.
 Finally,
  (\ref{TM}) and (\ref{dJ}) imply that $\ga=0$ for
 { $(\al_\ga,0,\ga)\in\ker Q_\rho\cap\aT_\om^0$.}
 Then also $\al_\ga=0$ by (\ref{Res202}).
 As a result, 
  $\ker Q_\rho\cap\aT_\om^0=0$.



\appendix

\protect\renewcommand{\thetheorem}{\Alph{section}.\arabic{theorem}}

\setcounter{equation}{0}
\section{On the soliton parameters}\la{aA}
Here we prove Lemma \ref{lmam}.

{\bf ad i).}
Substituting (\ref{solit3}) into  (\ref{pqv}), we obtain:
\beqn\nonumber
\pi_{v,\om}\!\!&\!\!=\!\!&\!\!\om-\!\langle \bA_{v,\om}(y)\we y\rho(y)\rangle
=I\om-\!\langle  \fr{\om \we\na\hat \rho(k)}{k^2\!-\!(v\cdot k)^2}\we\na\hat\rho(k)\rangle
=I\om-\!\int \fr{(\om \we k)\we k}{k^2-(v\cdot k)^2}\fr{|\na\rho(k)|^2\,dk}{k^2}\\
\la{PMvom3} 
\!\!&\!\!=\!\!&\!\!\om\Big(I+\!\int\fr{|\na\rho(k)|^2\, dk}{k^2-(v\cdot k)^2}\Big)-\!\int\fr{k(\om\cdot k)|\na\rho(k)|^2}{k^2(k^2-(v\cdot k)^2)} dk
=\om\Big(I+\!\int\fr{|\na\rho(k)|^2\, dk}{k^2-(v\cdot k)^2}\Big)-q_{v,\om},
\eeqn
where we denote $\check\rho_1(|k|):=\hat\rho(k)$. 
We consider the two cases $\om\Vert v$ and $\om\bot v$ separately.
\smallskip\\
{\bf The case $\om\Vert v$.}
We may assume that $\om=(|\om|,0,0)$ and $v=(\pm|v|,0,0)$. Then
\be\la{omwe}
q_{v,\om}:=\int\fr{k(\om\cdot k)|\na\rho(k)|^2}{k^2(k^2-(v\cdot k)^2)} dk=|\om|\int\fr{k_1k|\na\rho(k)|^2}{k^2(k^2-\!v^2 k_1^2)^2)} dk
=|\om| e_1\int\fr{k_1^2|\na\rho(k)|^2}{k^2(k^2-v^2k_1^2)} dk. 
\ee
Substituting into \eqref{PMvom3}, we obtain:
\be\la{omwe2}
\pi_{v,\om}=I_\eff^\Vert \om,\qquad I_\eff^\Vert=I_\eff^\Vert(v) =I+\langle  \fr{k_2^2+k_3^2}
{k^2(k^2\!-\!v^2 k_1^2)},|\na\rho(k)|^2\rangle.
\ee
{\bf The case $\om\bot v$.}
We may assume that $\om=(|\om|,0,0)$ and $v=(0,|v|,0)$.  In this case,
\be\la{omwe3}
q_{v,\om}=|\om| e_1\int\fr{k_1^2|\na\rho(k)|^2}{k^2(k^2-v^2k_2^2)} dk,\quad 
\pi_{v,\om}=I_\eff^\bot \om,\quad I_\eff^\bot=I+\langle  \fr{k_2^2+k_3^2}
{k^2(k^2\!-\!v^2 k_2^2)},|\na\rho(k)|^2\rangle
\ee
by \eqref{PMvom3}. Now Lemma \ref{lmam}  i) is proved.
\smallskip\\
{\bf ad ii).} Now we consider the case $(v,\om)\not\in \Si$. We may assume that  $v=(|v|,0,0)$ with $|v|\ne 0$ and  $\om=(\om_1,\om_2,0)$, where $\om_1\ne 0$ and $\om_2\ne 0$. In this case,
\beqn\nonumber
q_{v,\om}=
\int\begin{pmatrix}k_1^2\om_1\\
k_2^2\om_2\\0
\end{pmatrix}\frac{|\na\rho(k)|^2 dk}{k^2(k^2-v^2k_1^2)}=\begin{pmatrix}\om_1\alpha_1\\
\om_2\alpha_2\\0
\end{pmatrix}, \quad \alpha_j=\int\frac{k_j^2|\na\rho(k)|^2 dk}{k^2(k^2-v^2k_1^2)},~~j=1,2.
\eeqn
It remains to show that 
$\alpha_1>\alpha_2$ since then $q_{v,\om}\nparallel\om$,
so (\ref{piom2}) breaks down by (\ref{PMvom3}). Calculating in spherical coordinates, we obtain:
$$
\al_1=\frac{2\pi C_{\rho}}{|v|^3}\Big(\ln\frac{1+\!|v|}{1-\!v}-2|v|\Big),\qquad
\al_2=\frac{\pi C_{\rho}}{|v|^3}\Big((v^2-1)\ln\frac{1+\!|v|}{1-\!|v|}+2|v|\Big),
$$
where $C_{\rho}>0$ due to the last condition of (\ref{rosym}).
Hence, for $|v|<1$, we obtain:
$$
\al_1-\al_2=\frac{\pi C_{\rho}}{|v|^3}\Big((3-v^2)\ln\frac{1+|v|}{1-|v|}-6|v|\Big)
=\pi C_{\rho}\sum\limits_{k=2}^{\infty}\frac{8(k-1)}{(2k+1)(2k-1)}v^{2(k-2)}>0.
$$
\br 
The above inequality  and the formulas {\rm (\ref{omwe2})-(\ref{omwe3})}
 imply that $I_\eff^\bot >I_\eff^\Vert$ for $\rho(x)\not\equiv 0$.
\er

\setcounter{equation}{0}
\section{Density of $C_0^\infty$ in the space of divergence-free vector fields}
\la{aC}

Here we prove that $C_0^\infty$ is dense in the Hilbert space $\cF^0\!:=\!\{A\!\in\! L^2\!: \dv A(x)\!\equiv\! 0\}$. For any $A\in\cF^0$,
its Fourier transform satisfies $\hat A(k)\bot k$ for $k\in\R^3$. Hence,
$\hat A(k)=k\we \hat a(k)$ with $\hat a(k)=-\fr{k\we \hat A(k)}{k^2}$. For $m\in\N$,
let us denote
$$
 \hat B_m(k)=\left\{\ba{ll}\hat A(k),&|k|>\fr1m\\0,&|k|<\fr1m\ea\right|, \qquad
 \hat b_m(k)=-\fr{k\we \hat B_m(k)}{k^2}.
$$
Then 
\be\la{FAm}
\hat B_m(k)=k\we \hat b_m(k)\toLd \hat A(k),\qquad m\to\infty.
\ee
Note that $(1+|k|)\hat b_m(k)\in L^2$.
Hence,
$b_m\in H^1$, so there exist $b_{m,n}\in C_0^\infty$ such that $b_{m,n}\toH b_m$ as $n\to\infty$ for every $m$.
Therefore, 
$$
A_{m,n}:=\curl b_{m,n}\toLd \curl b_m,\quad n\to\infty\qquad {\rm and}\quad
\curl b_m=B_m\toLd A,\quad m\to\infty.
$$
Hence,
there exists a subsequence $A_{m,n(m)}\toLd A$ as $m\to\infty$. It remains to note that
$A_{m,n}\in C_0^\infty$ and  $\dv A_{m,n}(x)\equiv 0$.

\end{document}